\documentclass[aps,prl,twocolumn]{revtex4-1}
\usepackage{graphicx,subfigure}
\usepackage{amsmath,amssymb}
\usepackage{multirow}
\usepackage{textcomp}
\usepackage{float}
\usepackage{color}
\usepackage{pdfpages}
\newcommand{\avg}[1]{\langle #1 \rangle}
\newcommand{\Htwist}{\ensuremath{H_\textrm{twist}}}

\newcommand{\ket}[1]{\ensuremath{\vert {#1} \rangle}}
\newcommand{\bra}[1]{\ensuremath{\langle {#1} \vert}}
\newcommand{\vc}[1]{\ensuremath{\mathbf{#1}}}
\newcommand{\abs}[1]{\ensuremath{\left\vert #1 \right\vert}}

\newcommand{\up}{\ensuremath{\ket{\uparrow}}}
\newcommand{\down}{\ensuremath{\ket{\downarrow}}}
\newcommand{\exc}{\ensuremath{\ket{e}}}

\newcommand{\uvec}[1]{\ensuremath{\hat{\mathbf{#1}}}}

\renewcommand{\imath}{\ensuremath{\mathrm{i}}}

\newcommand{\var}[1]{\ensuremath{\left(\Delta {#1} \right)^2}}
\newcommand{\psiin}{\ensuremath{\ket{\uvec{x}}}}
\newcommand{\bpsiin}{\ensuremath{\bra{\uvec{x}}}}
\newcommand{\psie}{\ensuremath{\ket{\psi_e}}}

\newcommand{\rdet}{\ensuremath{\rho}}

\newcommand{\delt}{\ensuremath{d}}
\newcommand{\Gsc}{\ensuremath{\Gamma_\mathrm{sc}}}
\newcommand{\pflip}{\ensuremath{r}}

\DeclareMathOperator{\arccot}{arccot}

\newcommand{\tC}{\ensuremath{\overline{C}_6}}
\newcommand{\deltaC}{\ensuremath{\delta_{\scriptscriptstyle{C}}}}
\newcommand{\deltaR}{\ensuremath{\delta_{\scriptscriptstyle{R}}}}
\newcommand{\dSymeas}{\ensuremath{\Delta S_{\mathrm{meas}}}}
\newcommand{\dSCSS}{\ensuremath{\Delta S_{\mathrm{CSS}}}}

\usepackage{ulem}

\newcommand{\editout}[1]{{\bf \sout{#1}}}      
\renewcommand{\editout}[1]{}

\begin{document}
\title{Approaching the Heisenberg limit without single-particle detection}
\author{Emily Davis}
\affiliation{Department of Physics, Stanford University, Stanford, California 94305, USA}
\author{Gregory Bentsen}
\affiliation{Department of Physics, Stanford University, Stanford, California 94305, USA}
\author{Monika Schleier-Smith}
\affiliation{Department of Physics, Stanford University, Stanford, California 94305, USA}

\begin{abstract}
We propose an approach to quantum phase estimation that can attain precision near the Heisenberg limit without requiring single-particle-resolved state detection.  We show that the ``one-axis twisting'' interaction, well known for generating spin squeezing in atomic ensembles, can also amplify the output signal of an entanglement-enhanced interferometer to facilitate readout.  Applying this interaction-based readout to oversqueezed, non-Gaussian states yields a Heisenberg scaling in phase sensitivity, which persists in the presence of detection noise as large as the quantum projection noise of an unentangled ensemble.  Even in dissipative implementations---e.g., employing light-mediated interactions in an optical cavity or Rydberg dressing---the method significantly relaxes the detection resolution required for spectroscopy beyond the standard quantum limit.

\end{abstract}
\date{\today}

\maketitle

For decades, advances in atomic spectroscopy have brought clocks, accelerometers, and magnetometers to ever greater precision.  A recent development is the use of many-particle entangled states to reduce the statistical uncertainty in measurements of the energy difference $\hbar\omega$ between two atomic states $\up,\down$ \cite{Meyer01,Leroux10a,Louchet10,Gross10,Ockeloen13,Sewell12,Hamley12,Berrada13,Strobel14,Luecke11}.  Whereas an uncorrelated ensemble of $N$ two-level atoms achieves at best the standard quantum limit of precision $\Delta\omega T= 1/\sqrt{N}$ in an interrogation time $T$, entanglement can enhance this precision up to the fundamental Heisenberg limit $\Delta\omega T = 1/N$.  Approaching the Heisenberg limit with more than a few particles remains a major outstanding challenge, due to difficulties not only of preparing but also of detecting entangled quantum states \cite{Zhang12,Hume13,Bohnet14}.

Imperfect state detection has limited the sensitivity of entanglement-enhanced metrology with squeezed \cite{Meyer01,Louchet10,Gross10,Leroux10a,Sewell12,Hamley12,Ockeloen13,Berrada13}, oversqueezed \cite{Strobel14}, and twin Fock \cite{Luecke11} states.  The standard detection protocol is to measure the population difference $n\equiv n_\uparrow-n_\downarrow$ between the levels $\up,\down$ in the entangled state after perturbing it by an amount proportional to the frequency $\omega$, with $dn/d\omega = N T$.  Any uncertainty $\Delta n$ in the population measurement limits the attainable spectroscopic sensitivity to $\Delta\omega T > \Delta n/N$.  Correspondingly, approaching the Heisenberg limit requires single-particle-resolved state detection, which becomes increasingly difficult at large atom number.  Recent experiments have made progress in addressing this challenge \cite{Zhang12,Hume13}, but not yet under the conditions required to generate highly entangled states.

Theoretically, a quantum-enhanced measurement does not require directly detecting the entangled sensor state.  Several proposals instead envision echo protocols \cite{Yurke86,Toscano06,Goldstein11} in which a quantum system undergoes a unitary evolution $U$ into a non-classical state and, after subjecting this state to a perturbation, one attempts to reverse the evolution to the initial state by application of $U^\dagger$.  This approach in principle permits Heisenberg-limited measurements with an ensemble of spin-1/2 particles (or, equivalently, two-level atoms) by detecting the state of a single ancilla spin \cite{Goldstein11}.

In this Letter, we propose an echo protocol that enables spectroscopy near the Heisenberg limit with low-resolution state detection $\Delta{n}\sim \sqrt{N}$.  To generate entanglement, our method employs the global Ising interactions of the ``one-axis twisting'' Hamiltonian \cite{Kitagawa93}, realizable with cold atoms \cite{Sorensen01,Gross10,SchleierSmith10a,Leroux10,Zhang14,Ockeloen13,Sorensen02}, trapped ions \cite{Sorensen99,Monz11,Britton12} and solid-state nuclear spins \cite{Rudner11}.  Switching the sign of the interaction after subjecting the system to a weak perturbation amplifies the perturbation into a larger spin rotation that is easily detected.  We analyze the performance including dissipation in two atomic implementations, employing interactions mediated either by light in an optical cavity \cite{SchleierSmith10a,Zhang14} or by Rydberg dressing \cite{Gil14}.  In each case, the twisting echo enables precision far beyond the standard quantum limit with detection noise larger than the quantum noise of an unentangled state.

\begin{figure}[h]
\includegraphics[width=\columnwidth]{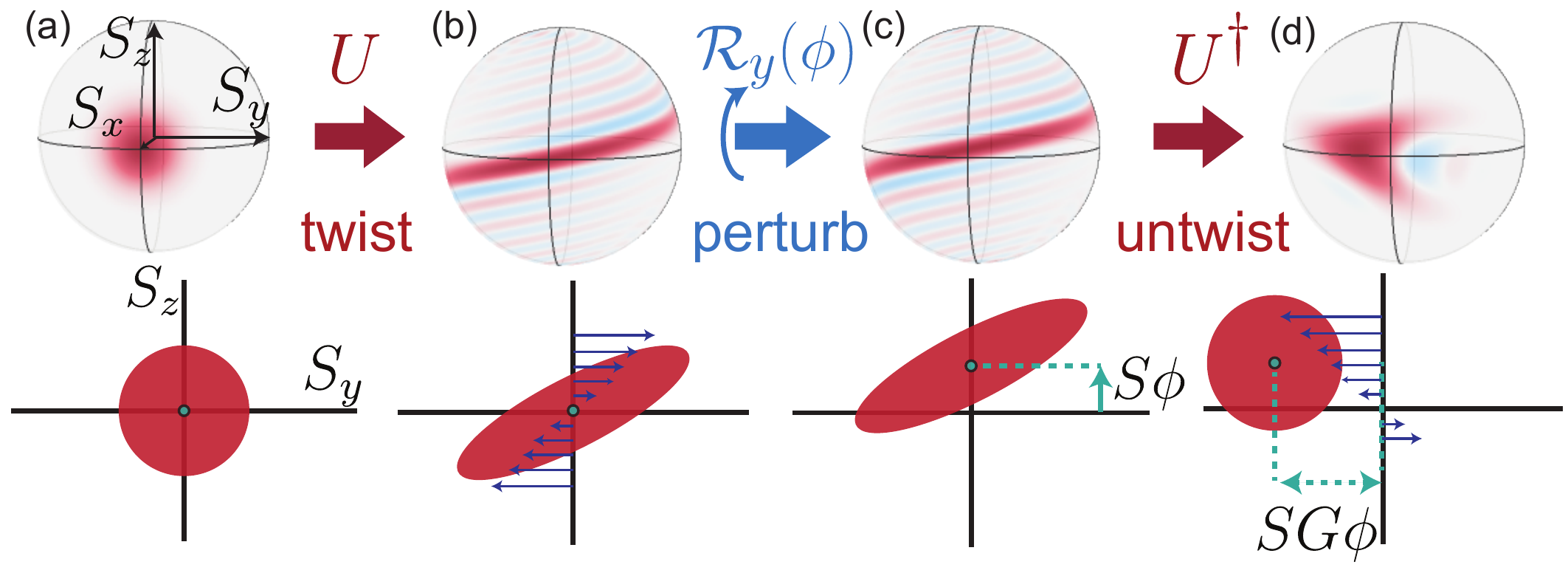}
\caption{\label{fig:echo_dynamics}{\bf Twisting echo} for entanglement-enhanced measurement.  Top row: The initial CSS $\psiin$ ({\bf a}) evolves under $\Htwist(\chi)$ into an oversqueezed state $\psie$ ({\bf b}).  To detect a rotation of $\psie$ about $\uvec{y}$ by a small angle $\phi$ (\textbf{b}$\rightarrow$\textbf{c}), we amplify the perturbation into a large displacement $\avg{S_y^\phi} = G S\phi$ by applying $\Htwist(-\chi)$ (\textbf{c}$\rightarrow$\textbf{d}).  Illustrated are Wigner quasiprobability distributions for $2S = 30$ atoms, with $\phi=1/S$. Bottom row: Cartoon depiction of the same steps, with blue flow lines indicating twisting and untwisting.}
\end{figure}

The one-axis twisting Hamiltonian $\Htwist = \chi S_z^2$ describes internal-state-dependent interactions in a collection of $N$ two-level atoms, which we represent in terms of spin-1/2 operators $\vc{s}_i$ by a collective spin $\vc{S}=\sum_{i=1}^N \vc{s}_i$.  The dynamical effect of $\Htwist$ is to generate a spin precession about the $\uvec{z}$-axis at a rate proportional to $S_z$ (Fig. \ref{fig:echo_dynamics}\textbf{a}-\textbf{b}).  For spins initially polarized along $\uvec{x}$, the lowest-order effect of $H_\mathrm{twist}$ is squeezing \cite{Kitagawa93}.  At longer interaction time, $H_\mathrm{twist}$ produces non-Gaussian states, including oversqueezed states and ultimately a maximally entangled GHZ state at $\chi t = \pi/2$ \cite{Monz11}.  While the GHZ state enables Heisenberg-limited measurements in few-particle systems with highly coherent interactions, we will show that the twisting echo protocol attains a Heisenberg scaling $\Delta\omega T\propto 1/N$ at significantly shorter evolution time.

The squeezed and oversqueezed states generated by $H_\mathrm{twist}$ are highly sensitive to spin rotations $\mathcal{R}_y(\phi) = e^{-i \phi S_y}$ about the $\uvec{y}$ axis.  Indicative of this sensitivity are the increased quantum fluctuations $\Delta S_y$ in Fig. \ref{fig:echo_dynamics}\textbf{b}, which lower the quantum Cram\'{e}r-Rao bound on the uncertainty $\Delta\phi \ge 1/(2\Delta S_y)$ \cite{Giovannetti11}.  We therefore present a protocol for measuring $\mathcal{R}_y(\phi)$, bearing in mind that a compound sequence $\mathcal{R}_y(\phi) = \mathcal{R}_x(-\pi/2) \mathcal{R}_z(\phi) \mathcal{R}_x(\pi/2)$ then allows for measuring a precession $\phi = \omega T$ about $\uvec{z}$, much as in Ramsey spectroscopy with squeezed states \cite{Meyer01,Louchet10,Gross10,Leroux10a,Ockeloen13}.

The twisting echo protocol is shown in Fig. \ref{fig:echo_dynamics}, where we assume unitary dynamics.  An ensemble is initialized in the coherent spin state (CSS) $\psiin$ satisfying $S_x\psiin = S\psiin$ (Fig. \ref{fig:echo_dynamics}\textbf{a}).  Applying $\Htwist(\chi) = \chi S_z^2$ for a time $t$ yields the entangled state $\psie=U\psiin$, where $U=e^{-i\chi S_z^2 t}$ (Fig. \ref{fig:echo_dynamics}\textbf{b}).  To detect a rotation $\psie\rightarrow\mathcal{R}_y(\phi)\psie$ by a small angle $\phi$, we attempt to undo the twisting by applying $\Htwist(-\chi)$.  For $\phi = 0$, the final state $U^\dagger\mathcal{R}_y(\phi) U\psiin$ is identical to the original CSS.  However, a non-zero angle $\phi$ (Fig. \ref{fig:echo_dynamics}\textbf{c}) biases the $S_z$-dependent spin precession to produce a large final value of $\avg{S_y}$ (Fig. \ref{fig:echo_dynamics}\textbf{d}).  Measuring $S_y$, by rotating the state and then detecting the population difference $n_\uparrow-n_\downarrow$, provides a sensitive estimate of $\phi$.

\begin{figure}[h]
\includegraphics[width=\columnwidth]{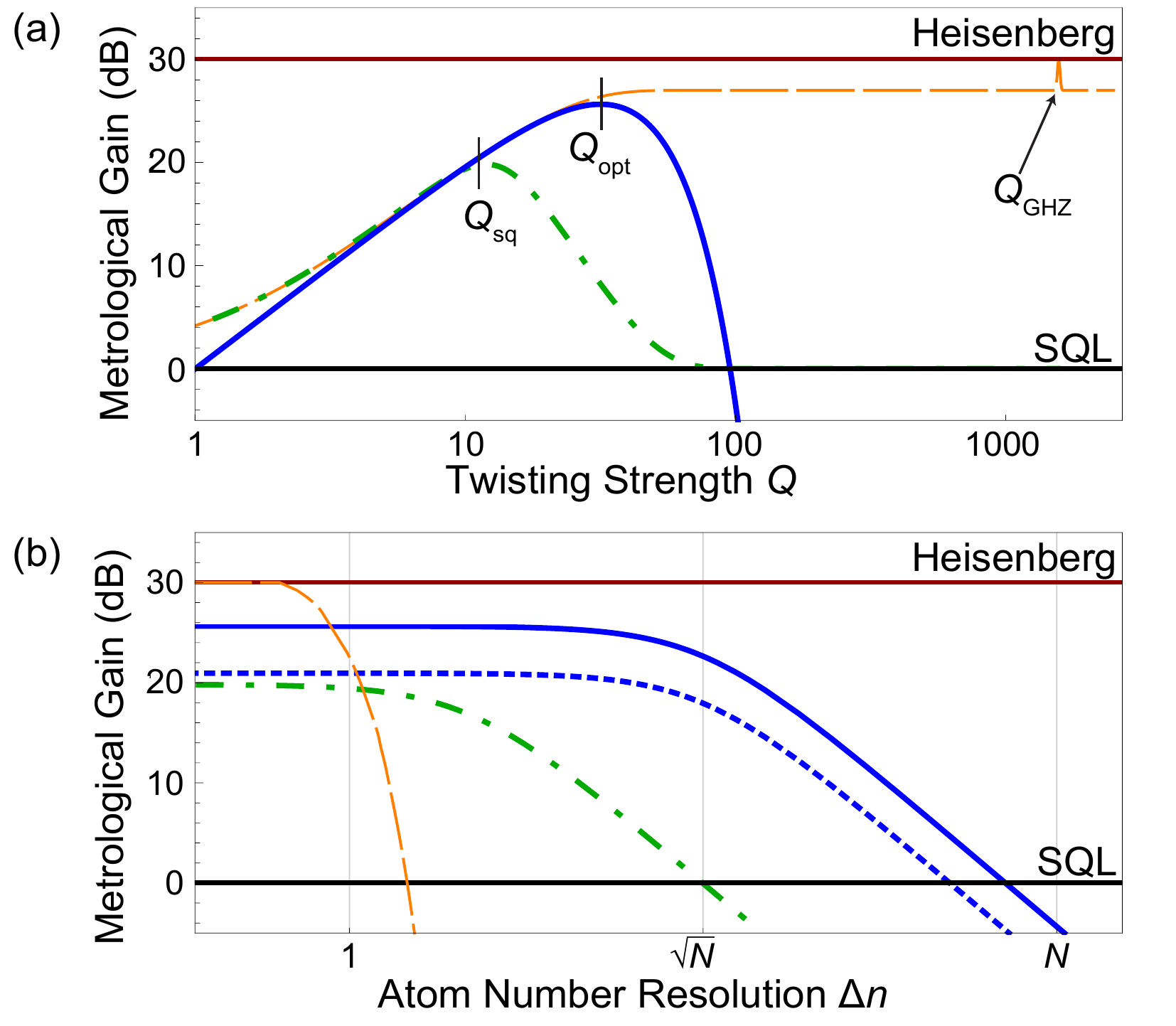}
\caption{\label{fig:echo_performance}{\bf Metrological gain of the twisting echo} with $N=10^3$ atoms.  Horizontal lines indicate the standard quantum limit (black) and Heisenberg limit (red). ({\bf a}) Metrological gain \textit{vs.} twisting strength $Q$ for the unitary twisting echo (solid blue), compared to spin squeezing (dot-dashed green) and the quantum Cram\'{e}r-Rao bound (QCRB) on phase sensitivity (dashed orange). The twisting echo nearly follows the QCRB to its plateau at $Q \approx \sqrt{N}$; only at a much longer time $Q_\mathrm{GHZ}=N\pi/2$ does the QCRB increase by 3 dB to reach the Heisenberg limit.  ({\bf b}) Metrological gain \textit{vs.} measurement uncertainty $\Delta n$ for echo with twisting strength $Q_\mathrm{opt}$ (solid blue) or $Q_{\mathrm{sq}}$ (dotted blue), compared to direct detection of the squeezed state at $Q_{\mathrm{sq}}$ (dot-dashed green). Dashed orange curve shows Cram\'{e}r-Rao bound for estimating $\phi$ in the  GHZ state $\mathcal{R}_y(\phi)(\ket{\uvec{y}} + \ket{-\uvec{y}})$ using projective measurements with uncertainty $\Delta n$.}
\end{figure}

The angular sensitivity is given by
\begin{equation}\label{eq:phase_sensitivity}
\Delta \phi = \left[\Delta S_y^\phi/\partial_\phi \avg{S_y^\phi}\right]_{\phi=0},
\end{equation}
where $\avg{S_y^{\phi}}$ and $\Delta S_y^{\phi}$ represent the mean and standard deviation of $S_y$ after the echo, and $\partial_\phi \equiv d/d\phi$.  The standard deviation for no rotation is ideally that of the initial CSS, $\Delta S_y^{\phi=0}=\Delta S_{\mathrm{CSS}}=\sqrt{S/2}$.  To evaluate the denominator of Eq. \ref{eq:phase_sensitivity}, we expand
\begin{align} \label{eq:calc}
\avg{S_y^{\phi}} &= \bpsiin U^{\dagger}e^{-i\phi S_{y}}US_{y}U^{\dagger}e^{i\phi S_{y}}U\psiin\nonumber\\
&= i\phi\bpsiin[S_{y},U^{\dagger}S_{y}U]\psiin + O(\phi^2).
\end{align}
to lowest order in $\phi$.  We express $S_y = (S_+ - S_-)/(2i)$ in terms of raising and lowering operators $S_\pm$ and simplify
\begin{equation}
U^\dagger S_\pm U = e^{-i\chi t S_z^2} S_\pm e^{i\chi t S_z^2} = S_\pm e^{i\chi t (\pm 2S_z+1)}
\end{equation}
to evaluate Eq. \ref{eq:calc} using the generating functions in Ref. \cite{Arecchi72}.  We thus arrive at a dependence
\begin{equation}\label{eq:dSydphi}
\left[\partial_\phi \avg{S_y^\phi}\right]_{\phi=0}= S(2S-1)\sin\left(\frac{Q}{2S}\right)\cos^{2S-2}\left(\frac{Q}{2S}\right)
\end{equation}
of the final spin orientation on the perturbation $\phi$, where we have introduced the `twisting strength' $Q \equiv 2 S \chi t$.  The resulting metrological gain $1/[N\var{\phi}]$ is plotted in Fig. \ref{fig:echo_performance}\textbf{a} as a function of $Q$ for $N=10^3$ atoms.  At the optimal twisting strength $Q_{\text{opt}} = 2 S \arccot(\sqrt{2S-2})\approx \sqrt{N}$ for $N\gg 1$, the echo protocol yields an angular sensitivity
\begin{equation}\label{eq:sqrte_over_N}
\Delta\phi_\mathrm{min} = \sqrt{e}/N.
\end{equation}
This sensitivity is very near the Heisenberg limit, despite a $\sim\sqrt{N}$-times shorter twisting evolution $Q_{\text{opt}}$ than required to reach a GHZ state ($Q_{\text{GHZ}}$ in Fig. \ref{fig:echo_performance}\textbf{a}).  The entangled state $\psie$ at $Q_{\text{opt}}$ is oversqueezed (Fig. \ref{fig:echo_dynamics}\textbf{b}), allowing the echo to surpass the sensitivity $\Delta\phi \propto 1/N^{5/6}$ attainable by spin squeezing \cite{Kitagawa93} under $H_\mathrm{twist}$.

The twisting echo is highly robust against detection noise (Fig. \ref{fig:echo_performance}\textbf{b}), as the ``untwisting'' amplifies the spin rotation signal by a factor of $G \equiv d\langle S_y^\phi\rangle/d(S\phi) \le \sqrt{N}$ (Fig. \ref{fig:echo_dynamics}\textbf{c}-\textbf{d}).  Concomitantly, the quantum noise returns to the CSS level, so that adding Gaussian detection noise  $\dSymeas=\rdet \dSCSS$ results in an angular sensitivity $\Delta\phi = \sqrt{1+\rdet^2}\Delta\phi_\mathrm{min}$. Thus, even a measurement that barely resolves a CSS, with atom number resolution $\Delta n = 2\Delta S_{\mathrm{meas}} \approx \sqrt{N}$, permits a sensitivity near the Heisenberg limit.  By contrast, measurement noise significantly degrades the sensitivity attainable by direct detection of non-Gaussian states: already at single-atom resolution, the twisting echo outperforms direct detection of a GHZ state (Fig. \ref{fig:echo_performance}\textbf{b}).

In practice, the sensitivity $\Delta\phi$ may be degraded by imperfect coherence of the one-axis twisting evolution.  To show that the twisting echo can provide a significant benefit in realistic metrological scenarios, we analyze the limitations due to dissipation in two implementations designed to enhance atomic clocks: the method of cavity feedback dynamics \cite{SchleierSmith10a} demonstrated in Refs. \cite{Leroux10,Leroux10a}; and the Rydberg dressing scheme proposed in Ref. \cite{Gil14}.

\begin{figure}[h]
	\includegraphics[width=\columnwidth]{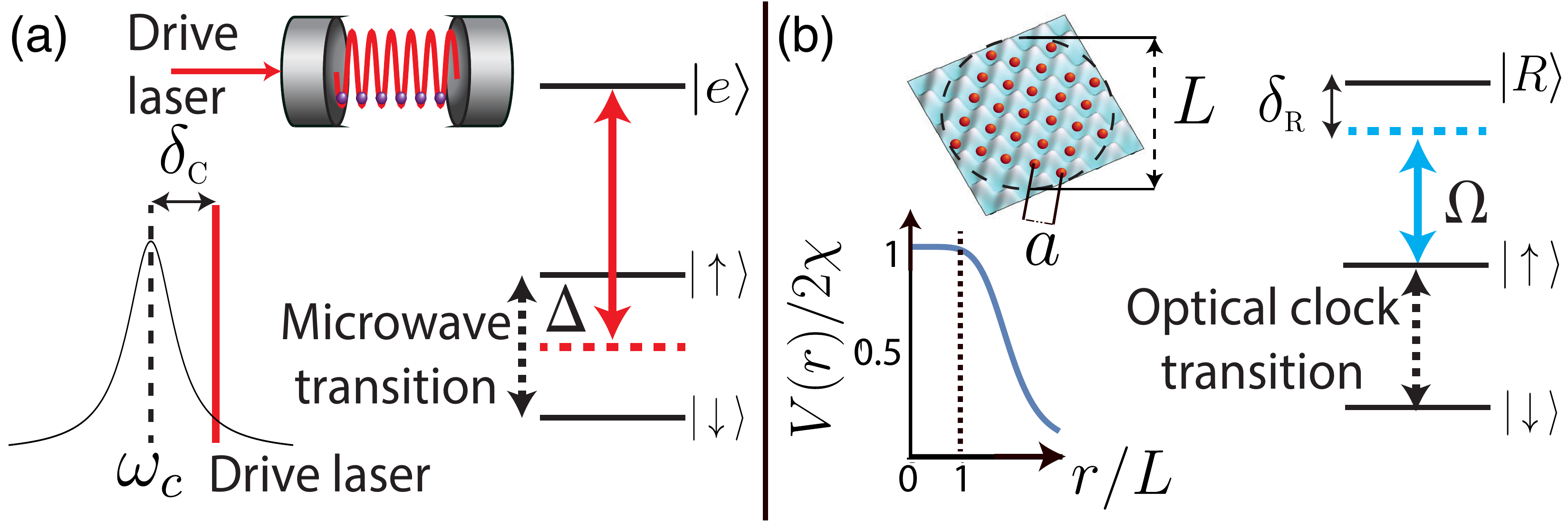}
	\caption{\label{fig:experimental_setups}{\bf Experimental schemes} for entanglement-enhanced measurement. ({\bf a}) Optical cavity with drive laser at detuning $\deltaC$ from cavity resonance $\omega_{c}$ and detunings $\pm\Delta$ from atomic transitions $\up\rightarrow\exc$, $\down\rightarrow\exc$.  ({\bf b}) Optical-lattice clock with metastable state $\up$ coupled to a Rydberg state $\ket{R}$ by a laser with Rabi frequency $\Omega$, producing the two-body potential $V(r)$ \cite{Gil14}.  Both schemes yield one-axis twisting dynamics, with the sign of $\Htwist$ dictated by the sign of $\deltaC$ ($\deltaR$) in the cavity-QED (Rydberg) system.}
\end{figure}

\textit{Cavity-Mediated Interactions ---} The scheme for one-axis twisting by light-mediated interactions \cite{SchleierSmith10a,Leroux10,Zhang14} is shown in Fig. \ref{fig:experimental_setups}\textbf{a}.  Atoms in hyperfine states $\up, \down$ are coupled to an optical resonator mode with vacuum Rabi frequency $2g$, at large detunings $\pm \Delta$ from transitions to an excited state $\exc$.  The dispersive atom-light interaction shifts the cavity resonance frequency in proportion to $S_z$, with $\partial \omega_c/\partial S_z = \Phi\kappa/2$, where $\Phi \equiv 4g^2/\Delta \kappa$.  Thus, driving the cavity at a detuning $\deltaC$ from the bare-cavity resonance results in an $S_z$-dependent intracavity power.  The latter acts back on the atomic levels via the a.c. Stark shift, yielding an $S_z$-dependent spin precession.  For small cavity shifts $\sqrt{N}\Phi\kappa\ll \deltaC$, the spin precession rate depends linearly on $S_z$, yielding one-axis twisting dynamics.  The sign of the twisting is controlled by $\deltaC$, while the strength depends on the average number of photons $p$ in the coherent field incident on the cavity: $Q = 2 S p \Phi^2 \delt/(1+\delt^2)^2$, where  $\delt \equiv 2\deltaC/\kappa$.

The light-induced twisting is accompanied by fluctuations in the phase of the collective spin due to photon shot noise.  These fluctuations are described by a Lindblad operator $L = \sqrt{\gamma}S_z$ in the master equation for the density matrix of the spin subsystem, where $\gamma = 2\chi/\delt$ \cite{SM}\nocite{Reiter}. Physically, $\gamma t$ represents the average number of photons rescattered into the cavity per atom while twisting.  The leakage of these photons from the cavity in principle enables a measurement of $S_z$, whose backaction is the dephasing described by a decay in $\avg{(S_\pm^{\phi=0})^2} = e^{-4\gamma t} \avg{(S_\pm^{t=0,\phi=0})^2}$ after twisting and untwisting.   Accordingly, to lowest order in $\gamma t \ll 1$, for $N\gg 1$, the variance of $S_y$ grows to
\begin{equation}\label{eq:Syvar2}
\var{S_y^{\phi=0}}/\Delta{S_\mathrm{CSS}}^2 \approx  1 + 4S\gamma t = 1 + 4 Q/d.
\end{equation}
Thus, cavity decay increases $\Delta\phi$ by a factor $\sqrt{1+4Q/d}$ compared with the unitary case.

The phase broadening can be made arbitarily small at large detuning $\deltaC$ at the price of increased decoherence from spontaneous emission.  The latter occurs at a rate $\Gsc = \chi(1/\delt+\delt)/(2\eta)$ per atom, where $\eta \equiv 4g^2/(\kappa\Gamma)$ is the single-atom cooperativity and $\Gamma$ is the excited-state linewidth.  Assuming each spontaneous emission event has a probability $\pflip$ of flipping a spin, the average value of $S_z$ while ``untwisting'' differs from that during ``twisting'' by a root-mean-square value \cite{SchleierSmith10a} $\Delta S_z^\mathrm{sc}\approx \sqrt{4\pflip S\Gsc t/3} = \sqrt{r Q(1/d+d)/(3\eta)}$.  Such a change  has the same effect on the final signal as a perturbation $\Delta\phi_\mathrm{sc} = \Delta S_z^\mathrm{sc}/S$, and thus contributes to the uncertainty in measuring $\phi$.

\begin{figure}[h]
\includegraphics[width=\columnwidth]{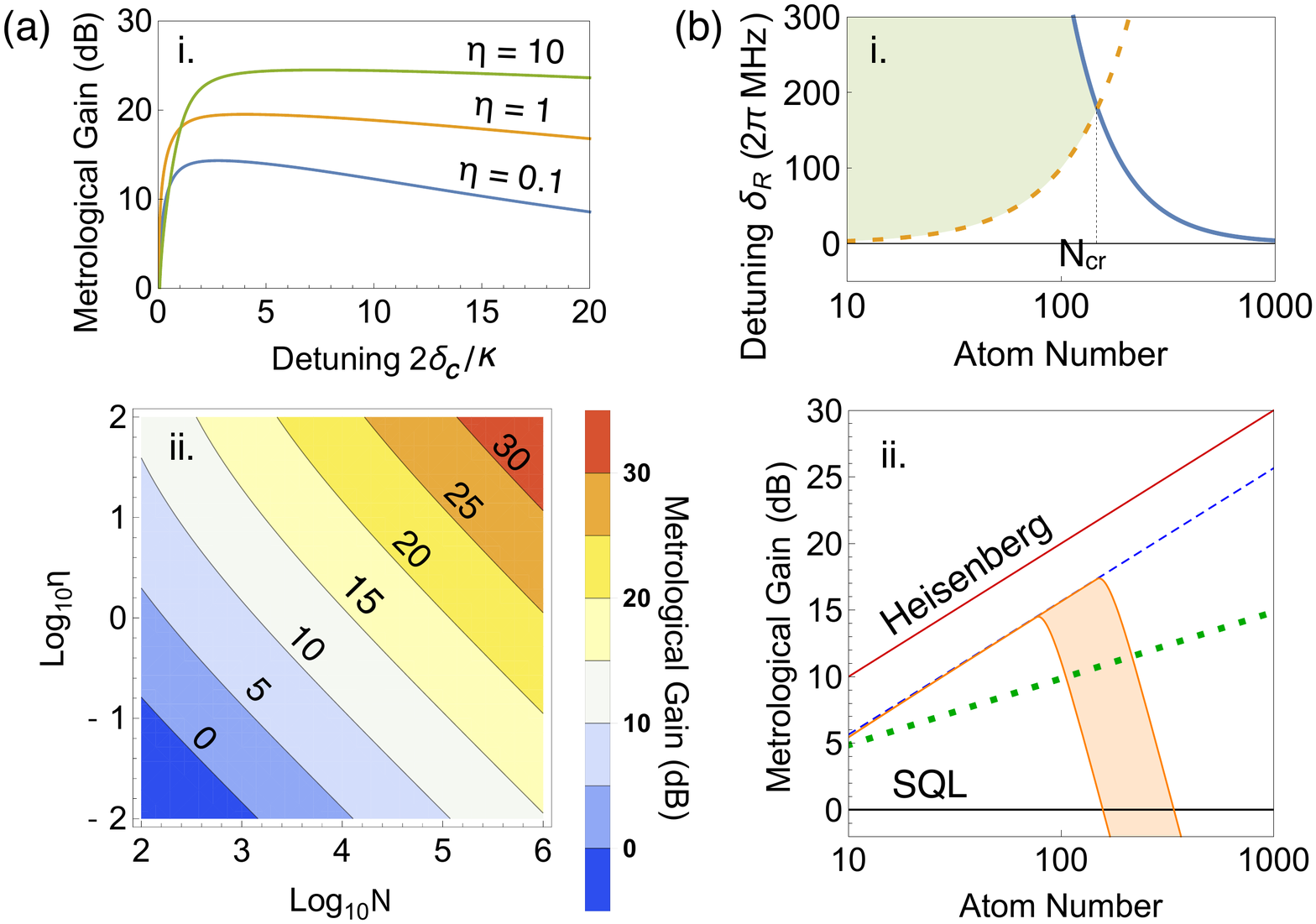}
\caption{\label{fig:echo_results}{\bf Dissipative twisting echo.}  ({\bf a}) Cavity-mediated twisting, with $r=1/2$. ({\bf i}) Metrological gain $-10\log_{10}\sigma^{2}$ \textit{vs.} laser-cavity detuning $\delt=2\deltaC/\kappa$ for $N=10^5$ atoms and cooperativity $\eta=0.1$ (blue), $1$ (orange),  $10$ (green). ({\bf ii}) Metrological gain \textit{vs.} $N$ and $\eta$. ({\bf b}) Rydberg-mediated twisting in Sr for $n=70$, $2a = 800$ nm, $\epsilon = 0.1$. ({\bf i}) Limitations on the detuning $\deltaR$, which must lie below the blue line for all atoms to fit inside the interaction sphere, but above the dotted orange line to avoid spontaneous emission. ({\bf ii}) Metrological gain  \textit{vs.} $N$ in Rydberg scheme limited by spontaneous emission (orange), compared to idealized case without spontaneous emission (dashed blue) and cavity-mediated twisting with $\eta = 10$ (dotted green). Orange band represents a range of $\tC$ coefficients $10^{10}\le \tC\le 10^{11}$ calculated for principal quantum numbers $60\lesssim n \lesssim 70$ \cite{Vaillant12,*Kunze93}.}
\end{figure}

To calculate the phase sensitivity of the dissipative echo, we first express the normalized phase variance $\sigma^2_0 \equiv 2S\var{\phi}_0$ for the unitary case in terms of the twisting strength $Q$.  From Eq. \ref{eq:dSydphi}, using the approximation $\cos^{2S-2}(\chi t) \approx e^{-S(\chi t)^2}$ for $\chi t\ll 1$, we obtain $\sigma^2_0 \approx e^{Q^2/(2S)}/Q^2$.  The total phase variance including cavity decay and spontaneous emission is then given by $\sigma^2 \equiv 2S\var{\phi} = \sigma^2_0 + \sigma^2_\mathrm{diss}$, where
\begin{equation}
\sigma^2_\mathrm{diss} = 4e^{Q^2/(2S)}/(Qd)+ 2rQ\left(\delt^{-1} + \delt \right)/(3S\eta)
\end{equation}
represents the noise added by dissipation.  This noise will reduce the optimal twisting strength below $Q_{\text{opt}} = \sqrt{2S}$, so that $e^{Q^2/(2S)}\sim 1$.  The dissipative contribution $\sigma^2_\mathrm{diss}$ is then minimized by choosing $Qd \approx \sqrt{6S\eta/r}$ and large detuning $d\gg 1$, yielding a total variance
\begin{equation}
\sigma^2 \approx r(1+ d^2)/(6S\eta) + \sqrt{32r(1+1/d^2)/(3S\eta)}.
\end{equation}
At large collective cooperativity $2S\eta \gg 1$, the normalized phase variance $\sigma^2$ depends only weakly on detuning (Fig. \ref{fig:echo_results}\textbf{a}.i) for ${1\ll d\ll \sqrt[4]{S\eta/r}}$, where we obtain $\sigma^2 \approx \sigma^2_\mathrm{diss}\approx \sqrt{32r/(3S\eta)}$.  We plot the metrological gain $\sigma^{-2}$ as a function of $N$ and $\eta$ in Fig. \ref{fig:echo_results}\textbf{a}.ii.

The scaling of the metrological gain with collective cooperativity is just as in spin squeezing by quantum non-demolition measurement \cite{Hammerer04}, but no coherence-preserving measurement is required.  Moreover, the measurement only requires a resolution on the order of the width $\Delta S^\phi_{y} \geq \dSCSS$ of the broadened final state. The benefits of the echo for state detection can be combined with schemes for reducing dissipation \cite{Saffman09,Bohnet14,Trail10,*Leroux12} to achieve even higher sensitivity.

\textit{Rydberg Dressing---}To approach the ideal unitary echo, we consider implementing $\Htwist$ by Rydberg dressing (Fig. \ref{fig:experimental_setups}\textbf{b}) in a strontium optical-lattice clock \cite{Gil14}.  Here, the pseudo-spin states are the singlet ground state $\down=\ket{^1 S_0}$ and metastable triplet state $\up=\ket{^3 P_0}$.  A laser with Rabi frequency $\Omega$ is detuned by $\deltaR$ from the $\up\rightarrow\ket{R}$ transition, where $\ket{R}=\ket{n ^3 S_1}$ represents a Rydberg state of principal quantum number $n$. The coupling to $\ket{R}$ induces a two-body potential $V(r)$ that is nearly constant over a distance $L = \frac{1}{2}[C_6 / (2 \deltaR)]^{1/6}$ (Fig. \ref{fig:experimental_setups}\textbf{b}), enabling all-to-all interactions among $N$ atoms confined within a region of diameter $L$.  In the limit of weak dressing, where the probability $\epsilon\equiv N_\uparrow \Omega^2/(4\deltaR^2)$ for even a single one of the $N_\uparrow\approx N/2$ spin-up atoms to be excited to $\ket{R}$ is small ($\epsilon \ll 1$), the result is a one-axis twisting Hamiltonian with interaction strength $\chi = \Omega^4/(16 \deltaR^3)$ \cite{Gil14}. The sign of the interaction is controlled by the detuning $\deltaR$, while the strength $\abs{\chi} = \epsilon^2 C_6/(2^5 N^2 L^6)$ is highest for small, dense ensembles.

Obtaining maximally coherent all-to-all interactions requires careful choice of the detuning from the Rydberg state.  The detuning $\deltaR$ controls both the interaction range $L$ and the ratio $\chi/\Gsc \propto \deltaR$ between the spin-spin coupling and the spontaneous emission rate $\Gsc = \epsilon\Gamma/N$ per atom, where $\Gamma$ is the Rydberg state linewidth.  To fit all atoms inside a sphere of diameter $L$ in a three-dimensional lattice with spacing $a$, we must restrict the detuning to $\deltaR < \Gamma \tC / (2^9 N^2)$, where $\tC \equiv C_6/(\Gamma a^6)$.  To conservatively estimate the metrological gain attainable under this restriction, without modeling the effects of decay to other Rydberg states, we restrict the twisting-untwisting evolution to the first spontaneous emission event: $2N\Gsc t\lesssim 1$.  Reaching the optimum oversqueezed state at $Q_\text{opt}$ then requires a detuning $\deltaR > \Gamma N^{3/2} / (2 \epsilon)$.  Both conditions on $\deltaR$, plotted in Fig. \ref{fig:echo_results}\textbf{b}.i, can be met simultaneously for up to $N_{\text{cr}} = (\tC \epsilon / 2^8)^{2/7}$ atoms. Here, $\tC \sim 10^{11}$ for a Rydberg state of principal quantum number $n \sim 70$ \cite{Vaillant12,*Kunze93} in a lattice of the ``magic'' wavelength $2a \approx 800~\text{nm}$ for the clock transition  \cite{Ye08,*Katori11}.  At a Rydberg-state population $\epsilon = 0.1$, the ideal phase sensitivity of Eq. \ref{eq:sqrte_over_N} can then be reached with up to $N_{\text{cr}}\approx 150$ atoms.

We compare the predicted performance of the Rydberg and cavity schemes in Fig. \ref{fig:echo_results}\textbf{b}.ii. For low atom number, the Rydberg dressing outperforms cavity-mediated interactions even at strong coupling $\eta \sim 10$.  Yet whereas the cavity echo improves monotonically with $N$, the Rydberg echo reaches an optimum at the critical atom number $N_{\text{cr}}$, above which the coherent evolution time must be reduced to extend the interaction range.  Even with only $N_{\text{cr}}\approx 150$ atoms, the twisting echo matches the phase sensitivity of $\sim 10^4$ unentangled atoms.  The method could thus benefit atomic clocks employing asynchronous interrogation of many small sub-ensembles \cite{Biedermann13,Borregaard13}.

We have presented a protocol that amplifies the effect of a phase rotation on an entangled state to enhance signal readout. By transferring the phase information to the average displacement of a near-Gaussian state, the twisting echo attains a Heisenberg scaling in sensitivity without single-particle resolution, and eliminates the need for Bayesian estimation methods for non-Gaussian states. Our approach can guide the design of new experiments by alleviating the need to simultaneously optimize coherence of interactions and fidelity of state detection. The protocol is adaptable to a wide range of systems, including ones where the sign of the interaction is fixed.  For example, spin rotations can convert one-axis twisting to a two-axis counter-twisting Hamiltonian $H_\mathrm{ct} \propto S_{x}^{2}-S_{y}^{2}$ \cite{Liu11} and can switch the sign of $H_\mathrm{ct}$ to exchange the squeezed and amplified quadratures. Future work might explore extensions to richer many-body systems featuring finite-range interactions or chaotic dynamics.

\begin{acknowledgments}This work was supported by the AFOSR, the Alfred P. Sloan Foundation, the NSF, and the Fannie and John Hertz Foundation.  We thank V. Vuleti\'{c}, A.~S. S{\o}rensen, M. Kasevich, and O. Hosten for helpful discussions.
\end{acknowledgments}


\bibliography{twisting_echo_refs_arxiv}
\clearpage


\section*{Supplementary material}

# Approaching the Heisenberg Limit Without Single-Particle Detection: Supplemental Material

In this supplement, we derive the effective one-axis twisting dynamics of an atomic ensemble coupled to a driven optical cavity [1–3]. This system has been analyzed elsewhere, but we know of no comprehensive treatment using the Lindblad master equation formalism. We also establish an equivalence between the twisting dynamics of the driven cavity scheme and the twisting dynamics of a four-photon Raman scheme proposed by Sørensen and Mølmer [4].

## PHYSICAL SYSTEM

The model system consists of $N$ three-level atoms coupled to single cavity mode with annihilation operator $a$. The total Hamiltonian is:

$$H = H_c + H_a + H_{\text{int}} + H_{\text{drive}} \tag{S1}$$

where $H_c, H_a$ are the cavity and atomic Hamiltonians, respectively:

$$H_c := \omega_c a^\dagger a$$
$$H_a := \sum_{i=1}^{N} \left[ 2\Delta \left|\uparrow\right\rangle \left\langle\uparrow\right|_i + (\omega_c + \Delta) \left|e\right\rangle \left\langle e\right|_i \right]. \tag{S2}$$

The two atomic ground-state levels $\left|\downarrow\right\rangle_i, \left|\uparrow\right\rangle_i$ are hyperfine sublevels separated by an energy difference $2\Delta$ and the excited state $\left|e\right\rangle_i$ is accessible from the ground states by optical transitions (see Fig. 3a in the main text). The cavity frequency $\omega_c$ is detuned an equal amount $\pm\Delta$ from the ground states $\left|\downarrow\right\rangle_i, \left|\uparrow\right\rangle_i$.

The dipole interaction term $H_{\text{int}}$ describes the coupling between the atoms and the cavity,

$$H_{\text{int}} := \sum_{i=1}^{N} \left[ ga \left|e\right\rangle \left\langle\uparrow\right|_i + ga \left|e\right\rangle \left\langle\downarrow\right|_i + \text{h.c.} \right], \tag{S3}$$

which excites the atom to $\left|e\right\rangle_i$ by annihilating a cavity photon with Rabi frequency $2g$. Finally, the drive term $H_{\text{drive}}$ populates the cavity with photons from an external coherent driving field $\beta$ at frequency $\omega_l$:

$$H_{\text{drive}} := \sqrt{\frac{\kappa}{2}} \left[ \beta^* a \, e^{-i\omega_l t} + \beta a^\dagger \, e^{i\omega_l t} \right], \tag{S4}$$

where $\kappa$ is the cavity linewidth. In general, the drive laser is detuned from cavity resonance by an amount $\delta_c := \omega_l - \omega_c$.

Including dissipative processes, the dynamics are described by a Lindblad master equation for the density matrix $\rho$:

$$\dot\rho = i\left[H, \rho\right] + \frac{1}{2} \sum_j \left( \left[L_j \rho, L_j^\dagger\right] + \left[L_j, \rho L_j^\dagger\right] \right), \tag{S5}$$

where Lindblad relaxation operators $L_j$ describe the various decay processes. We introduce a relaxation operator $L$ to describe the loss of photons from the cavity,

$$L = \sqrt{\kappa} \, a, \tag{S6}$$

and $2N$ relaxation operators for the spontaneous emission processes,

$$L_{i,\sigma} = \sqrt{\Gamma} \left|\sigma\right\rangle \left\langle e\right|_i, \tag{S7}$$

where $\Gamma$ is the spontaneous emission rate and $\sigma = \uparrow, \downarrow$ labels the atomic state after emission.

We are interested in the effective dynamics of the ground states $\left|\downarrow\right\rangle_i, \left|\uparrow\right\rangle_i$. This is a large Hilbert space, but we can restrict the dynamics to a much smaller subspace by noting that the Hamiltonian couples the atoms to the cavity *symmetrically*. Therefore the atomic ensemble can simply be viewed as a single spin-$N/2$ that lives in the $(N+1)$-dimensional symmetric Dicke subspace. Unfortunately, the spontaneous emission operators $L_{i,\sigma}$ break this symmetry,

so in this analysis we will ignore them by assuming that the twisting dynamics occur before we lose a significant fraction of photons through emission. (The limitations imposed by spontaneous emission on the twisting evolution are analyzed heuristically in the main text.)

The goal is to derive the effective dynamics of this spin-$N/2$ in the presence of the driven cavity. We do this by adiabatically eliminating first the excited states $|e\rangle_i$ and then the cavity $a$ from the dynamics, leaving an effective Hamiltonian describing only the dynamics of the spin. To accomplish this, we follow the procedure of Reiter and Sorenson [5]. This procedure relies on two assumptions:

1. The dynamics of the ground states $|\downarrow\rangle_i, |\uparrow\rangle_i$ are slow compared to the dynamics of the rest of the system, and

2. The ground states $|\downarrow\rangle_i, |\uparrow\rangle_i$ are only perturbatively affected by the fast dynamics of the rest of the system.

The first assumption allows us to solve for the fast dynamics in the presence of essentially stationary ground states, while the second ensures that the effective ground state dynamics *remain* slow even when coupled to the rest of the system. We refer the reader to the original paper for the full discussion.

## ADIABATIC ELIMINATION OF $|e\rangle_i$

We first eliminate the excited states $|e\rangle_i$ using the Reiter-Sorenson procedure. To do so we must make two assumptions:

1. $\sqrt{\kappa}|\beta|\sqrt{\langle a^\dagger a\rangle} \ll \Delta$ (i.e. excited state dynamics are fast compared to other timescales)

2. $g\sqrt{\langle a^\dagger a\rangle} \ll \Delta$ (i.e. coupling to the excited states is perturbative)

The second assumption is necessary in order to ignore the ground-state driving terms $\sim \sqrt{\frac{\kappa}{2}}\beta a^\dagger \, e^{-i\delta_c t}$ compared to the ground-state splitting $2\Delta$.

Ignoring overall energy shifts, the result of this procedure is:

$$H_{\text{eff}} = (\omega_c + \Phi\kappa S_z/2)\, a^\dagger a + 2\Delta S_z + \sqrt{\frac{\kappa}{2}}\left[\beta^* a\, e^{i\omega_l t} + \beta a^\dagger\, e^{-i\omega_l t}\right]$$
$$L_{\text{eff}} = \sqrt{\kappa}\, a \tag{S8}$$

where $\Phi\kappa := 4g^2/\Delta$ (assuming $\Gamma \ll \Delta$) and $S_z := \frac{1}{2}\sum_i \sigma_z^i = \frac{1}{2}\sum_i [|\uparrow\rangle\langle\uparrow|_i - |\downarrow\rangle\langle\downarrow|_i]$. We see that the effect of the atomic ensemble is to shift the cavity resonance by $\Phi\kappa S_z/2$. In particular, we note that this cavity shift depends on the $z$-projection $S_z$ of the ensemble spin.

Passing to a rotating frame with respect to:

$$H_0 = 2\Delta S_z + \omega_l a^\dagger a \tag{S9}$$

we obtain a time-independent effective Hamiltonian:

$$H'_{\text{eff}} = (-\delta_c + \Phi\kappa S_z/2)\, a^\dagger a + \sqrt{\frac{\kappa}{2}}\left[\beta^* a + \beta a^\dagger\right]$$
$$L'_{\text{eff}} = \sqrt{\kappa}\, a \tag{S10}$$

## MACROSCOPIC POPULATION OF CAVITY

Due to the drive laser, the cavity is highly populated with photons, so we are not justified in simply adiabatically eliminating the occupied cavity mode $a$. Formally, the existence of a strong drive laser and populated cavity breaks the second assumption required for adiabatic elimination because the coupling terms $\sim \sqrt{\frac{\kappa}{2}}\beta^* a$ are not perturbative.

However, we can always decompose the cavity field $a$ into a large classical field $\alpha$ accompanied by small quantum fluctuations $c$:

$$a = \alpha + c. \tag{S11}$$

(Alternatively, this can be viewed as a displacement of the cavity operator $a$ by an amount $\alpha$.) We take the classical field $\alpha$ to be the steady-state population that would be induced by the driving field $\beta$ in the absence of atoms:

$$\alpha = \beta \sqrt{\frac{\kappa}{2}} \frac{1}{\delta_c + i\kappa/2} \tag{S12}$$

Then the operator $c$ describes the fluctuations in the cavity field due to its interaction with the atoms, which can be treated perturbatively. Plugging Eq. S12 into $H'_{\text{eff}}$ and $L'_{\text{eff}}$ we obtain:

$$\begin{aligned} H''_{\text{eff}} &= (-\delta_c + \Phi\kappa S_z/2) \, c^\dagger c + \alpha^* \Phi\kappa S_z/2 \, c + \alpha \Phi\kappa S_z/2 \, c^\dagger + \Phi\kappa S_z/2 \, |\alpha|^2 \\ L''_{\text{eff}} &= \sqrt{\kappa} \, c \end{aligned} \tag{S13}$$

where we have redefined the Hamiltonian to absorb the constant term that appears in $L'_{\text{eff}}$.

## ADIABATIC ELIMINATION OF CAVITY FLUCTUATIONS

We now eliminate the cavity fluctuations $c$. Applying the Reiter-Sorenson procedure once again, we obtain the result:

$$\begin{aligned} H_{\text{spin}} &= -\frac{-\delta_c + \Phi\kappa S_z/2}{(-\delta_c + \Phi\kappa S_z/2)^2 + \kappa^2/4} \, \Phi^2 \kappa^2 |\alpha|^2 /4 \, S_z^2 \\ L_{\text{spin}} &= \sqrt{\frac{\kappa}{(-\delta_c + \Phi\kappa S_z/2)^2 + \kappa^2/4}} \, \Phi\kappa |\alpha|/2 \, S_z \end{aligned} \tag{S14}$$

where we have made the following assumption:

3. $\Phi\kappa S_z |\alpha|/2 \ll |\delta_c| + \Phi\kappa S_z/2$ (i.e. coupling to excited states is perturbative)

The spin Hamiltonian simplifies further if in addition we assume that the shift $\Phi\kappa S_z/2$ in cavity frequency is small:

4. $\Phi\kappa S_z/2 \ll |\delta_c|$

In this case we obtain:

$$\begin{aligned} H_{\text{spin}} &= \chi \, S_z^2 \\ L_{\text{spin}} &= \sqrt{2\chi/d} \, S_z \end{aligned} \tag{S15}$$

where we have defined:

$$\begin{aligned} d &:= 2\delta_c/\kappa \\ \chi &:= \frac{d}{(1+d^2)^2} \frac{dp}{dt} \Phi^2 \end{aligned} \tag{S16}$$

and $dp/dt = |\beta|^2$ is the rate at which coherent photons are incident on the cavity. During this analysis, we ignored the effect of spontaneous emission. We can, however, relate the rate of spontaneously emitted photons per atom to the interaction strength $\chi$ as:

$$\Gamma_{\text{sc}} = \left(\frac{g}{\Delta}\right)^2 |\alpha|^2 \Gamma = \frac{|\chi|}{2\eta} \frac{1+d^2}{d} \tag{S17}$$

where $\eta := 4g^2/(\kappa\Gamma)$ is the single-atom cooperativity. We use Eq. S17 to estimate the effects of spontaneous emission in the main text. (While this estimate does not treat the dynamical effects of spontaneous emission, it does give a sense of the timescale $\sim 1/(N\Gamma_{\text{sc}})$ over which we can reasonably ignore these effects.)

We reiterate the four assumptions that were made during this derivation, written in terms of the parameters $d, \Phi, dp/dt$:

1. $\frac{dp}{dt}\sqrt{\frac{2}{1+d^2}} \ll \Delta$

2. $\sqrt{\frac{\Phi}{2}\frac{dp/dt}{1+d^2}} \ll \sqrt{\Delta}$

3. $\sqrt{\frac{dp/dt}{\kappa}\frac{2}{1+d^2}}\,\Phi S_z \ll d$ (using assumption 4 to ignore $\Phi S_z$)

4. $\Phi S_z \ll d$

In our system, where the atoms are initialized in a coherent spin state in the $x$–$y$ plane, assumption 4 is satisfied by keeping $\Phi\sqrt{N} \ll d$, and assumptions 1-3 can always be satisfied by reducing the rate $dp/dt$ of photons incident on the cavity.

As a final note, it turns out that the effective spin dynamics in equation (S15) are equivalent to the dynamics one obtains from another one-axis twisting proposal that at first glance appears to be quite different. Sørenson and Mølmer proposed a four-photon Raman scheme to generate a one-axis twisting Hamiltonian by directly driving an atomic ensemble (e.g., using beams transverse to the cavity); the atoms communicate with one another by exchanging photons via a weakly occupied cavity mode [4]. The dynamics of this scheme are described by the effective operators:

$$H_{\text{spin, SM}} = \xi\,S_x^2$$
$$L_{\text{spin, SM}} = \sqrt{2\xi/d}\,S_x \qquad (S18)$$

where we've defined

$$\xi := \frac{d}{1+d^2}\frac{\Omega_R^2}{\Delta}\Phi \qquad (S19)$$

and $\Omega_R$ is the Rabi frequency of the transverse drive lasers. In this scheme, the quantities $\Delta, \delta_c$ are the single-photon detuning and two-photon detuning, respectively (other parameters, such as the cavity coupling $g$, are common to both schemes). The rate of spontaneous emission per atom in this scheme is:

$$\Gamma_{\text{sc, SM}} = 2\left(\frac{\Omega_R}{2\Delta}\right)^2\Gamma = \frac{|\xi|}{2\eta}\frac{1+d^2}{d} \qquad (S20)$$

Comparing equation (S15) to (S18) and equation (S17) to (S20) we see that, up to a rotation of spin axes $S_x \to S_z$, the two schemes produce precisely the same dynamics.

Although the two schemes are equivalent, the Sørenson-Mølmer scheme may offer an advantage for phase detection in that it twists about the $S_x$ axis. Thus, in an echo sequence that begins with a CSS along $\hat{\mathbf{y}}$, the twisting generates a state that is sensitive to a rotation about $S_z$, which the untwisting amplifies into a large signal in $\langle S_z \rangle$ that can directly be detected in the populations of states $|\uparrow\rangle, |\downarrow\rangle$. By contrast, in the driven cavity scheme one must apply two extra rotations to perform an amplified phase measurement.

---




\end{document}